\documentclass[aps,prb,amsmath,amssymb,twocolumn,10pt,footinbib,showpacs]{revtex4-1}
\usepackage{graphicx}
\usepackage{graphics}
\usepackage{amsmath}
\usepackage{amssymb}
\usepackage{amsfonts}
\usepackage{dsfont}
\usepackage{sidecap}
\usepackage{hyperref}
\usepackage{color}

\makeatletter
\newcommand*{\rom}[1]{\expandafter\@slowromancap\romannumeral #1@}
\makeatother

\newcommand*\oline[1]{%
  \vbox{%
    \hrule height 0.5pt
    \kern0.25ex
    \hbox{%
      \kern-0.1em
      \ifmmode#1\else\ensuremath{#1}\fi
      \kern-0.1em
    }
  }
}

\begin{document}

\title{Composite Majorana Fermion Wavefunctions in Nanowires}
\author{Jelena Klinovaja}
\author{Daniel Loss}
\affiliation{Department of Physics, University of Basel, Klingelbergstrasse 82, 4056 Basel,
Switzerland}
\date{\today}
\pacs{73.63.Nm; 74.45.+c}

\begin{abstract}
We consider Majorana fermions (MFs) in quasi-one-dimensional nanowire systems containing normal and superconducting sections where the topological phase based on Rashba spin orbit interaction can be tuned by  magnetic fields. We derive explicit analytic solutions of the MF wavefunction in the weak and strong spin orbit interaction regimes. We find that the wavefunction for one single MF is a composite object formed by superpositions of different MF wavefunctions which have nearly disjoint supports in momentum space. 
These contributions are coming from the extrema of the spectrum, one  centered around zero momentum  and  the other around the two Fermi points. 
As a result, the various MF wavefunctions
have different localization lengths in real space and  interference among them leads to pronounced oscillations of the MF probability density. 
For a transparent normal-superconducting junction we find that in the topological phase the MF leaks out from the superconducting  into the normal section of the wire and is delocalized over the entire normal section, in agreement with recent numerical results by Chevallier {\it et al.}.

\end{abstract}

\maketitle
\section{Introduction} Majorana fermions\cite{Majorana} (MFs), being their own antiparticles,
have attracted much attention in recent years  in condensed matter physics \cite{kitaev, fu, Nagaosa_2009,Sato,Akhmerov, lutchyn_majorana_wire_2010, oreg_majorana_wire_2010, potter_majoranas_2011, alicea_majoranas_2010, Flensberg_2010,Qi, suhas_majorana, stoudenmire, lutchyn, Klinovaja_CNT, half_metal_Duckheim, aguado_1, alicea_review_2012}.
Besides being of fundamental interest, 
these exotic quantum particles  have the potential for being used in topological quantum computing due to their non-Abelian statistics\cite{QC-Kitaev, QC-Freedman, QC-Nayak, QC-alicea_NP_2011, QC-Bonderson, 
QC-Bravyi, QC-Bravyi-2}. There are a number of systems where to expect MFs, e.g. fractional quantum Hall systems\cite{Read_2000,Nayak}, topological insulators\cite{fu, Nagaosa_2009}, optical lattices\cite{Sato},  $p$-wave superconductors\cite{potter_majoranas_2011}, and 
especially nanowires with strong Rashba spin orbit interaction\cite{lutchyn_majorana_wire_2010, oreg_majorana_wire_2010, alicea_majoranas_2010} - the system of interest in this work. There are now several  claims for experimental evidence of MFs in topological insulators \cite{Sasaki_2011,Goldhaber_2012} and, in particular, in semiconducting nanowires of the type considered here\cite{delft_MF, sweden_MF, indiana_MF}.

As is well-known\cite{lutchyn_majorana_wire_2010, alicea_majoranas_2010, oreg_majorana_wire_2010, alicea_review_2012}, an $s$-wave superconductor brought into 
contact with a semiconducting nanowire with Rashba spin orbit interaction (SOI) 
induces effective $p$-wave superconductivity that gives rise to MFs,  one at each end of such a wire. 
Most studies  have analyzed the corresponding model Hamiltonian by direct numerical diagonalization, which provides  exact solutions of the Schr\"odinger equation for 
essentially all  parameter values irrespective of their relative sizes. 
Less attention, however, has been given to analytical approaches which can provide additional insights into the nature of MFs. 
As usual, this comes with a price: closed analytic expressions  are hard to come by and can be obtained  only in special limits. But since these limits turn out to include realistic parameter regimes  such an approach is not a mere academic exercise but  worthwhile also  from a physical point of view.

Motivated by this, we focus in the present work on the spinor-wavefunction for MFs, and derive analytical expressions for various limiting cases, loosely characterized as
weak and strong  SOI regimes.
We find  that these solutions are superpositions of states  that come, in general, from different extremal points of the energy dispersion, one centered
around  zero-momentum  and  the others around the Fermi points. Despite having nearly disjoint support in momentum-space, 
 all such contributions must be taken into account, in general, in order to satisfy the boundary conditions imposed on the spinor-wavefunctions in real space. As a consequence of this composite structure
 of the MF wavefunctions, there will be more than one localization length that characterizes a single MF.  We will see throughout this work that the Schr\"odinger equation for the systems under consideration allows, in principle, degenerate MF wavefunctions. However,
 this degeneracy gets completely removed by the boundary conditions considered here, and, consequently, there exists only one single   MF wavefunction at a given end of the nanowire.
The superposition also gives rise to interference effects that leads to pronounced oscillations of the MF probability density in real space. 
Quite interestingly, the relative strengths of the different localization lengths as well as of the oscillation periods can be tuned by magnetic fields.

If only a section of the wire is covered with a superconductor, a normal-superconducting (NS) junction is formed. For this case, we find that the MF becomes delocalized over the entire normal section, while still localized in the superconducting section, as noted by several groups before\cite{Yakovenko, alicea_review_2012, Sticlet_2012, Sau_2012, Cheng}, and most recently studied in detail in a numerical study by Chevallier {\it et al.} \cite{Chevallier}.
Here, we will find analytical solutions for this problem, valid in the weak and strong SOI regime.
Depending on the length of the normal section, the support of the MF wavefunction is, again, centered at zero momentum or the Fermi momenta. Also similarly as before,   different localization lengths and oscillation periods of the MF
in the normal section occur, again tunable by  magnetic fields. This could then provide an experimental signature for MFs, e.g. in a tunneling density of states measurement, where a signal that comes from  a zero-mode MF will show oscillations along the normal section.

The paper is organized as follows.  In Sec. \ref{section_model} we introduce the continuum model of a nanowire including SOI, magnetic field and induced superconductivity. The composite structure of MF in proximity-induced superconducting wire is discussed in Sec. \ref{section_MF_SC} for  strong and weak SOI. In Sec. \ref{section_normal_region} we investigate an NS junction and show how the type of MF wavefunction oscillates in space and  depends on  magnetic field.  The final Sec. \ref{section_conclusions} contains our conclusions. Some technical details are referred to two Appendices.

\section{Model \label{section_model}}

\begin{figure}[tbp]
    \centering
    \includegraphics[width=8cm]{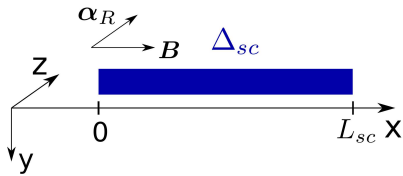}
\caption{
Nanowire (blue slab) of length $L_{sc}$ in the superconducting regime with  gap $\Delta_{sc}$ induced via proximity effect by a  bulk $s$-wave superconductor
(not shown).
A magnetic field $\bf B$ is applied along the nanowire in $x$-direction and perpendicularly to the Rashba SOI vector $\boldsymbol{\alpha}_R$ that points in $z$-direction.
}
\label{wire}
\end{figure}

Following earlier work\cite{lutchyn_majorana_wire_2010,oreg_majorana_wire_2010,alicea_majoranas_2010,Chevallier, alicea_review_2012}, our starting point is a semiconducting nanowire with Rashba SOI (see Fig. \ref{wire}) characterized by a SOI vector $\boldsymbol \alpha_R$ that points perpendicularly to the nanowire axis and defines the spin quantization direction $z$. In addition, a magnetic field ${\bf {B}}$ is applied along the nanowire in $x$-direction. We imagine that  the nanowire (or a section of it) is  in tunnel-contact with a conventional bulk
$s$-wave superconductor which leads to proximity-induced superconductivity in the nanowire itself, characterized by the induced $s$-wave gap $\Delta_{sc}$ (see Fig. \ref{wire}). We refer to this part of the nanowire as to the superconducting section (or as to the nanowire being in the superconducting regime), in contrast to the `normal' section of the nanowire that is not in contact with the superconductor
and thus in the normal regime.

We describe this nanowire system by a continuum model and our goal is to find the explicit wavefunctions for the MFs in the entire nanowire, including normal and superconducting section. 
For this, we need to introduce some basic definitions and briefly recall well-known results about the spectrum.

\begin{figure}[!b]
    \centering
    \includegraphics[width=8cm]{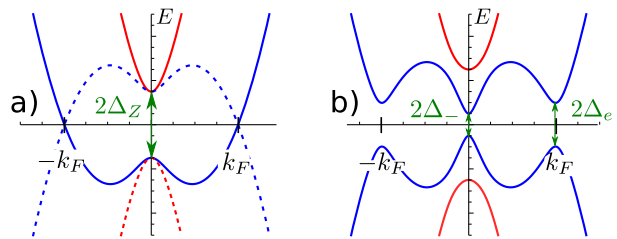}
\caption{Bulk spectrum for extended electron (solid lines) and hole (dashed lines) states in the normal (a) and in the superconducting regime (b). 
(a) In the normal regime, a Zeeman gap $2\Delta_Z$ is opened at $k=0$, but the full spectrum is still gapless due to the propagating modes at the Fermi points $\pm k_F$. (b) The proximity-induced superconductivity leads to the opening of a gap $\Delta_e$ at the Fermi points $\pm k_F$ and modifies the topological gap $\Delta_-=\Delta_{sc}-\Delta_Z$ at $k=0$.}
\label{spectrum}
\end{figure}

The Hamiltonian $H^0=H^{kin}+H^{SOI}+H^{Z}$ for the normal regime \cite{lutchyn_majorana_wire_2010,oreg_majorana_wire_2010} consists of the kinetic energy term
\begin{equation}
H^{kin}=\sum_{\sigma} \int dx\  \Psi_\sigma^\dagger (x)\left[\frac{(-i \hbar \partial_x)^2}{2m}-\mu\right]\Psi_{\sigma} (x),
\end{equation}
where $m$ is the (effective) electron mass and $\mu$ the chemical potential,
the SOI term,
\begin{equation}
H^{SOI}= -i \alpha_R \sum_{\sigma, \sigma'} \int dx \ \Psi_\sigma^\dagger (x)(\sigma_3)_{\sigma \sigma'} \partial_x \Psi_{\sigma'} (x),
\end{equation}
where, again, the  $z$-axis is chosen along $\boldsymbol \alpha_R$, and the Zeeman term corresponding to the magnetic field $B$ along the nanowire ($x$-axis),
\begin{equation}
H^{Z}=\Delta_Z \sum_{\sigma, \sigma'} \int dx \ \Psi_\sigma^\dagger (x)(\sigma_1)_{\sigma \sigma'}\Psi_{\sigma'} (x).
\end{equation}
Here, $\Psi^\dagger_\sigma(x)$ is the creation operator of an electron at position $x$ with spin $\sigma/2=\pm 1/2$ (along $z$-axis), and the Pauli matrices $\sigma_{1,2,3}$ act on the spin of the electron.
The Zeeman energy is given by $\Delta_Z= g \mu_B B/2$, where $g$ is the g-factor and $\mu_B$ the Bohr magneton.
It is convenient to introduce the corresponding Hamiltonian density $\mathcal{H}^0$,
\begin{align}
H^0 &= \int dx\ \psi^\dagger (x) \mathcal{H}^0 \psi (x),\nonumber\\
\mathcal{H}^0& =- \hbar^2 \partial_x ^2/2m-\mu - i \alpha_R \sigma_3  \partial_x +\Delta_Z  \sigma_1,
\label{h_0_density}
\end{align}
which acts on the vector $\psi = (\Psi_\uparrow, \Psi_\downarrow)$. The bulk spectrum of $\mathcal{H}^0$ (see Fig. \ref{spectrum}a) consists of two branches and is given by 
\begin{equation}
E_{\pm}^0(k) = \frac{\hbar^2 k^2}{2 m} - \mu \pm \sqrt{(\alpha_R k)^2 + \Delta_Z ^2}\,
,\label{spectrum-OSC}
\end{equation}
where $k$ is a momentum along the nanowire. 
 By opening a Zeeman gap $2\Delta_Z$, the magnetic field lifts the spin degeneracy at $k=0$. The chemical potential $\mu$ is tuned inside this gap and set to zero. In this case, the Fermi wavevector is determined from $E_{-}^0(k_F)=0$ and given by
\begin{equation}
k_F = \sqrt{ 2k_{so}^2 + \sqrt{4k_{so}^4 + k_Z^4}}\, ,
\label{eq_kF}
\end{equation}
where $k_{so}=m\alpha_R/\hbar^2$ and $k_Z = \sqrt{2 \Delta_Z m }/\hbar$.

The nanowire in the superconducting regime is described by the Hamiltonian $H^0+H^{sc}$, where the  $s$-wave BCS Hamiltonian $H^{sc}$ couples states with opposite momenta and spins\cite{lutchyn_majorana_wire_2010,oreg_majorana_wire_2010},
\begin{equation}
H^{sc}=\frac{1}{2}\sum_{\sigma, \sigma'}\int dx\ \Delta_{sc}( \Psi_\sigma(i\sigma_2)_{\sigma\sigma'} \Psi_{\sigma'}+h.c.).
\label{hm_sc}
\end{equation}
The proximity-induced superconductivity gap $\Delta_{sc}$ is chosen to be real (thereby assuming that we can neglect the flux induced by the B-field, which is the case e.g. for InSb nanowires \cite{delft_MF}).  The spectrum of $H^0+H^{sc}$ (see Fig. \ref{spectrum}b) is then found to be
\begin{align}
E^2_{\pm}(k)=&\left(\frac{\hbar^2 k^2}{2 m}\right)^2+(\alpha_R k)^2 + \Delta_Z ^2+\Delta_{sc}^2\\ \nonumber
&\pm2\sqrt{\Delta_Z ^2\Delta_{sc}^2 +\left(\frac{\hbar^2 k^2}{2 m}\right)^2 \left(\Delta_Z ^2 + (\alpha_R k)^2  \right)}.
\end{align}
The `topological' gap at $k=0$ is given by $\Delta_-= \Delta_{sc}-\Delta_Z$, and the closing of this gap marks the transition between nontopological  ($\Delta_- > 0$) and topological ($\Delta_- < 0$) phases \cite{Sato,lutchyn_majorana_wire_2010,oreg_majorana_wire_2010}. 
In contrast, the gap at  $k_F$, $\Delta_{e}\equiv2|E_-(k_F)|$, is always nonzero (see Fig. \ref{spectrum}b).

\section{Majorana fermions in the superconducting section \label{section_MF_SC}}

In this section we consider first the simpler case where the superconducting section extends over the entire nanowire  from $x=0$ to $x=L_{sc}$, see Fig. \ref{wire}. In the topological phase there is one MF bound state  at each end of the nanowire\cite{lutchyn_majorana_wire_2010,oreg_majorana_wire_2010}. In the physically interesting regime, these two MFs should be independent and have negligible spatial overlap. This justifies the consideration of a semi-infinite nanowire.  In this work we focus on the MF at the left end, $x=0$.

We will consider two limiting regimes, namely strong  ($k_F \simeq 2 k_{so}$) and weak ($k_F \simeq  k_{Z}$) SOI. In both regimes, the Hamiltonian can be linearized near the Fermi points and solved analytically. We show that the MF wavefunction has support in $k$-space from the exterior ($k\simeq \pm k_F$) and the interior  ($k\simeq 0$) branches of the spectrum, see Fig. \ref{Strong_SOI}a. If the system is in some intermediate regime of moderate SOI, the support of the MF wavefunction extends over all momenta from $-k_F$ to $k_F$, and this case cannot be treated analytically in the linearization approximation considered here.

\subsection{Regime of strong SOI and rotating frame \label{section_MF_SC_strong}}
The regime of strong SOI is defined by the condition that the SOI energy at the Fermi level is larger than the Zeeman splitting, $\Delta_Z \ll m \alpha_R^2/ \hbar^2 $ (or
$k_F\approx 2 k_{so}$), and larger than the proximity gap, $ \Delta_{sc} \ll m \alpha_R^2/ \hbar^2 $. This allows us to treat the magnetic field and the proximity-induced superconductivity as small perturbations.

\begin{figure}[tbp]
    \centering
    \includegraphics[width=8cm]{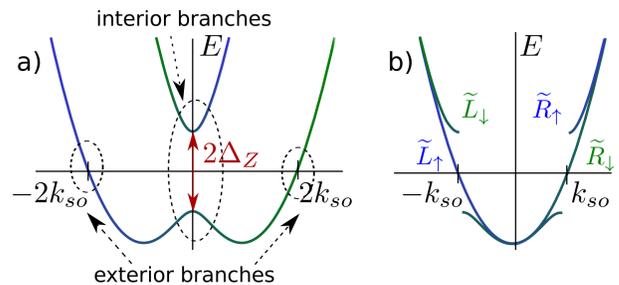}
\caption{ (a) Band structure of a nanowire with strong SOI and in a uniform magnetic field $B$ in the lab frame (see also Fig. \ref{spectrum}) for the normal section. States around $k=0$ belong to the interior branches and states around $k=k_F=2k_{so}$ belong to the exterior branches. (b) The same band structure in the rotating frame. The rotating magnetic field ${\bf  \widetilde{B}}(x)$ given by Eq. (\ref{mag_field}) couples $\widetilde{R}_\uparrow$ and $\widetilde{L}_\downarrow$ that leads to the opening of the Zeeman gap $2\Delta_Z$ but does not affect $\widetilde{L}_\uparrow$ and $\widetilde{R}_\downarrow$.}
\label{Strong_SOI}
\end{figure}

The spectrum  obtained in Eq. (\ref{spectrum-OSC}) consists of two parabolas shifted by the SOI momentum $k_{so}=m\alpha_R/\hbar^2$ and with a Zeeman gap opened at $k=0$ (see Fig. \ref{Strong_SOI}a). In the strong SOI regime it is more convenient to work in the rotating frame, see Fig. \ref{Strong_SOI}b. For this we follow Ref.~\onlinecite{braunecker} and make use of the following spin-dependent gauge transformation 
\begin{equation}
 \Psi_\sigma (x) = e^{-i\sigma k_{so} x }{\widetilde{\Psi} }_\sigma (x),
 \label{gaugetrafo}
\end{equation}
where tilde refers to the rotating frame. 
The $H^{SOI}$ term is effectively eliminated ($\widetilde{H}^{SOI}=0$)  
and the spectrum corresponding to $\widetilde{H}^{kin}$ consists of  
two parabolas centered at $k=0$, one for spin up and one for spin down.
Around the Fermi points, $\pm k_{so}$, the spectrum can be linearized and the electron operators $\widetilde{ \Psi}_\sigma$ are expressed in terms of slowly-varying right ($\widetilde{R}_\sigma$) - and left ($\widetilde{L}_\sigma$) - movers,
\begin{equation}
\widetilde{ \Psi}_\sigma (x) = \widetilde{R}_\sigma(x) e^{i k_{so} x}+\widetilde{ L}_\sigma(x) e^{-i k_{so} x}.
\label{operator}
\end{equation}
The kinetic energy term in the linearized model is 
\begin{align}
\widetilde{H}^{kin}
=-i \hbar \upsilon_F \int  dx \ [\widetilde{ R}_\sigma^\dagger (x)\partial_x \widetilde{R}_\sigma (x)-\ \widetilde{L}_\sigma^\dagger (x)\partial_x \widetilde{L}_\sigma (x)]
\end{align}
with Fermi velocity $\upsilon_F=\alpha_R/\hbar$. Here, we droped all fast oscillating terms, which is justified as long as $\xi\gg2\pi/k_{so}$, where $\xi$ is a localization length of $\widetilde{R}_\sigma$ and $\widetilde{L}_\sigma$ (see below). 

In the rotating frame the B-field becomes helical, rotating in the plane perpendicular to the SOI vector $\boldsymbol \alpha_R$,
\begin{equation}
{\bf  \widetilde{B}}(x)= B [\hat{x}\cos (2 k_{so} x)-\hat{y}\sin (2 k_{so} x)].
\label{mag_field}
\end{equation}
Here, $\hat{x}$ and $\hat{y}$ are unit vectors in $x$ and $y$ directions, respectively (see Fig. \ref{wire}). This leads to the Zeeman Hamiltonian of the form
\begin{align}
\widetilde{ H}^{Z}&=\Delta_Z  \int dx \ \widetilde{  \Psi}_\sigma^\dagger (x)e^{ 2 i \sigma  k_{so} x}\widetilde{  \Psi}_{-\sigma} (x),\nonumber\\
&\simeq\Delta_Z \int dx\ [\widetilde{R}_\uparrow^\dagger (x)\widetilde{L}_\downarrow (x)+\widetilde{L}^\dagger _\downarrow (x)\widetilde{R}_\uparrow(x)],
\end{align}
where in the second line we used the linearization approximation and, again, droped all fast oscillating terms. 
We note that only $\widetilde{R}_\uparrow (x)$ and $\widetilde{L}_\downarrow (x)$ are coupled, which leads to opening of a gap, as shown in Fig. \ref{Strong_SOI}b. This is similar to the spin-selective Peierls mechanism discovered in Ref. \onlinecite{braunecker} where interaction effects strongly renormalize this gap (here, however, we shall ignore interaction effects).

The superconductivity term [see Eq. (\ref{hm_sc})]  in the linearized model becomes
\begin{align}
\widetilde{H}^{sc}=&\frac{1}{2}\int dx\ \Delta_{sc} (\widetilde{R}_\uparrow (x)\widetilde{L}_\downarrow (x)-\widetilde{L}_\downarrow (x)\widetilde{R}_\uparrow(x) \nonumber \\
&+\widetilde{L}_\uparrow (x)\widetilde{R}_\downarrow(x)-\widetilde{R}_\downarrow (x)\widetilde{L}_\uparrow (x)+ h.c.).
\label{hm_sc_l}
\end{align}

We construct two vectors, $\widetilde{\phi}^{(i)}=(\widetilde{R}_\uparrow, \widetilde{L}_\downarrow, \widetilde{R}^\dagger_\uparrow, \widetilde{L}^\dagger_\downarrow)$ and $\widetilde{\phi}^{(e)}=(\widetilde{L}_\uparrow, \widetilde{R}_\downarrow, \widetilde{L}^\dagger_\uparrow, \widetilde{R}^\dagger_\downarrow)$, which correspond to the exterior ($k\simeq2k_{so}$) and interior  ($k\simeq0$)  branches of the spectrum in the lab frame (see Fig. \ref{Strong_SOI}a). 
The linearized Hamiltonian $\widetilde{ H}^{kin}+\widetilde{  
H}^{Z}+\widetilde{H}^{sc}$ reduces then to ($l=i,e$)
\begin{equation}
\widetilde{H}^{(l)} =\frac{1}{2}\int dx\ (\widetilde{\phi}^{(l)}(x))^\dagger \mathcal{\widetilde{H}}^{(l)} \widetilde{\phi}^{(l)}(x),
\label{h_lin}
\end{equation}
where the interior branches are described by
\begin{equation}
\mathcal{\widetilde{H}}^{(i)} =  -i \hbar \upsilon_F \sigma_3\partial_x  + \Delta_z \sigma_1 \eta_3 + \Delta_{sc} \sigma_2 \eta_2,
\label{H_int}
\end{equation}
and the exterior ones by
\begin{equation}
\mathcal{\widetilde{H}}^{(e)}=  i \hbar \upsilon_F \sigma_3 \partial_x  + \Delta_{sc} \sigma_2 \eta_2.
\label{H_ext}
\end{equation}
Here, 
the Pauli matrices $\eta_{1,2,3}$ act on the electron-hole subspace.

The energy spectrum is determined by the Schr\"{o}dinger equation, $\mathcal{\widetilde{H}}^{(l)} \widetilde{\varphi}^{(l)}_E (x)=E\widetilde{\varphi}^{(l)}_E (x)$, with boundary conditions to be imposed 
on the eigenfunctions $\widetilde{\varphi}^{(l)}_E (x)$ as discussed below.
We introduce then the operator 
$\widetilde{\gamma}^{(l)}_E=\int dx\ \widetilde{\varphi}_{E}^{(l)}(x)\cdot\widetilde{\phi}^{(l)}(x)$
and see that it diagonalizes Eq. (\ref{h_lin}), {\it i.e.}, $\widetilde{H}^{(l)}=\sum_E E (\widetilde{\gamma}^{(l)}_E)^\dagger \widetilde{\gamma}^{(l)}_E$. 
Focusing now on the zero modes, 
we consider in particular $\widetilde{\gamma}^{(l)}\equiv\widetilde{\gamma}^{(l)}_{E=0}$ but express it in a more convenient basis,
\begin{equation}
\widetilde{\gamma}^{(l)} = \int dx\ \widetilde{\varPhi}_{E=0}^{(l)}(x)\cdot\widetilde{\Psi}(x),
\label{Gamma_oper}
\end{equation}
where $\widetilde{\Psi} = (\widetilde{\Psi}_\uparrow, \widetilde{\Psi}_\downarrow, \widetilde{\Psi}_\uparrow^\dagger, \widetilde{\Psi}_\downarrow^\dagger)$ and where fast oscillating terms were dropped.
In this new basis $\widetilde{\Psi}$, we have reinstalled the phase factors  $e^{\pm i k_F x}$ (associated with $\widetilde{R}_{\sigma}$ and $\widetilde{L}_{\sigma}$) explicitly in 
the wavefunctions $\widetilde{\varPhi}_{E=0}^{(l)}(x)$, so that they are taken automatically into account when we impose the boundary conditions on $\widetilde{\varPhi}_{E=0}^{(l)}(x)$.

The zero-energy operator $\widetilde{\gamma}^{(l)}$ represents a MF, {\it i.e.},
$(\widetilde{\gamma}^{(l)})^\dagger=\widetilde{\gamma}^{(l)}$, 
if and only if the corresponding wavefunction  $\widetilde{\varPhi}_{E=0}^{(l)} (x)$ has the following form
\begin{equation}
 \widetilde{\varPhi}_{E=0}^{(l)} (x)=
\begin{pmatrix}
f(x)\\ g(x)\\ f^*(x)\\ g^*(x)                                      
\end{pmatrix},
\end{equation}
where the functions $f,g$
are arbitrary up to normalization $\int dx |\widetilde{\varPhi}_{E=0}^{(l)} (x)|^2/2=\int dx (|f(x)|^2+|g(x)|^2)=1$, which, however, will be suppressed in the following. 
 
In infinite space (no boundary conditions), the spectrum of the interior branches [see Eq. (\ref{H_int})] is given by $E_{\pm}^{(i)}=\pm\sqrt{(\hbar\upsilon_F \kappa)^2+\Delta_{\pm}^2}$,  
while the one for the 
exterior branches [see Eq. (\ref{H_ext})] is given by $E_{1,2}^{(e)}=\pm\sqrt{(\hbar\upsilon_F \kappa)^2+\Delta_{e}^2}$, where $\kappa$ is taken from the Fermi point.
Here, $\Delta_{\pm}=\Delta_{sc}\pm\Delta_{Z}$ and $\Delta_{e}=\Delta_{sc}$. If $\Delta_Z$ and $\Delta_{sc}$ become equal, the topological interior  gap $\Delta_-$ is closed. In contrast, the exterior gap $\Delta_e$ is not affected by the magnetic field.

The only normalizable  eigenstates of $\mathcal{\widetilde{H}}^{(l)}$ at zero energy and at $x>0$ are two evanescent modes coming from the interior branches, characterized by a decay wavevector $k^{(i)}_{\pm}=|\Delta_{\pm}|/\alpha_R$,  and two evanescent modes coming from the exterior branches, characterized by a decay wavevector $k^{(e)} =\Delta_{sc}/\alpha_R$. The corresponding zero-energy eigenfunctions $\widetilde{\varphi}_{E=0}^{(l)} (x)$ 
give the four basis wavefunctions, exponentially decaying in the semi-infinite space $x>0$,
\begin{widetext}
\begin{align}
\widetilde{\varPhi}^{(i)}_-=\begin{pmatrix}
-i\ {\rm sgn} (\Delta_{-}) e^{-ik_{so} x}\\
e^{ik_{so} x}\\
i\ {\rm sgn} (\Delta_{-})e^{ik_{so} x}\\
e^{-ik_{so} x}
\end{pmatrix} e^{-k_-^{(i)}x},\ \ \ \ \ \ \ \ \ \ \ \ \ \ \ \ \ \ 
\widetilde{\varPhi}^{(i)}_+=\begin{pmatrix}
 e^{-ik_{so} x}\\
-i\ e^{ik_{so} x}\\
 e^{ik_{so} x}\\
i\ e^{-ik_{so} x}
\end{pmatrix} e^{-k_+^{(i)} x}, \label{vectors_inter_1}
 \\[5pt]
\widetilde{ \varPhi}^{(e)}_1=\begin{pmatrix}
i\ e^{ik_{so} x}\\
e^{-ik_{so} x}\\
-i\ e^{-ik_{so} x}\\
e^{ik_{so} x}
\end{pmatrix} e^{-k^{(e)} x},\ \ \ \ \ \ \ \ \ \ \ \ \ \ \ \ \ \ \ 
\widetilde{ \varPhi}^{(e)}_2=\begin{pmatrix}
 e^{ik_{so} x}\\
i\ e^{-ik_{so} x}\\
 e^{-ik_{so} x}\\
-i\ e^{ik_{so} x}
\end{pmatrix} e^{-k^{(e)} x}.
\label{vectors_inter_2}
\end{align} 
\end{widetext}

Here we should note that these four degenerate  MF wavefunctions are not yet solutions of our problem: they do satisfy the Schr\"{o}dinger equation but not yet the boundary conditions. Thus, we search now for a linear combination of them, $\widetilde{ \varPhi}_M$, such that the boundary conditions are satisfied. At the left end of the nanowire, the  condition on the wavefunction is $\widetilde{\varPhi}_M(x=0)=0$. We assume here that the length of the nanowire $L_{sc}$ provides the largest scale, so we can neglect any interplay between the two ends of the nanowire and treat them independently (see also below).
The set of vectors $\{\widetilde{\varPhi}^{(i)}_-,\ \widetilde{\varPhi}^{(i)}_+,\ \widetilde{ \varPhi}^{(e)}_1,\ \widetilde{ \varPhi}^{(e)}_2\}$ is seen to be linearly independent in the nontopological phase at $x=0$, thus it is impossible to satisfy the boundary conditions and no solution exists at zero energy. In contrast, in the topological phase, $\Delta_Z>\Delta_{sc}$,  the two vectors $\widetilde{\varPhi}^{(i)}_-$ and $ \widetilde{
  \varPhi}^{(e)}_1$ are `collinear' such 
that the boundary condition can be satisfied and the zero energy state is a MF given by $\widetilde{\varPhi}_M=\widetilde{\varPhi}^{(i)}_--\widetilde{ \varPhi}^{(e)}_1$ in the rotating frame.
Using Eq. (\ref{gaugetrafo}), the MF wavefunction in the lab frame is then given by 
\begin{align}
\varPhi_{M}(x)=\begin{pmatrix}
i\\
1\\
-i\\
1
\end{pmatrix}   e^{-k_-^{(i)}x} -
\begin{pmatrix}
i\ e^{ik_F x}\\
 e^{-ik_F x}\\
-i\ e^{-ik_F x}\\
 e^{ik_F x}
\end{pmatrix}e^{-k^{(e)} x},\label{wave_MF}
\end{align}
with $k_F=2k_{so}$. 

There are a few remarks in order. First, we see that the initial fourfold degeneracy of the MF  has been completely removed by the boundary condition and we end up with one single non-degenerate MF wavefunction at the left end of the nanowire, $x=0$ (analogously for the right end, $x=L_{sc}$). 
This is reminiscent of a   well-known fact in elementary quantum mechanics, where for spinless particles in a one-dimensional box the degeneracy also gets removed by vanishing boundary conditions (whereas there is degeneracy for periodic boundary conditions).\cite{Landau_3} This non-degeneracy of the MFs is a generic feature which will occur in all cases considered in this work, even
in the presence of additional symmetries such as pseudo-time reversal invariance (see Sec. \ref{sec:sc_weak} below and footnote \onlinecite{footnote_time_reversal}).

Second, we see that the MF wavefunction $\varPhi_{M}$ is a `composite' object that is a superposition of two MF wavefunctions with (essentially) disjoint supports in $k$-space, one coming from the exterior ($\widetilde{ \varPhi}^{(e)}_1$)  and one from interior ($\widetilde{\varPhi}^{(i)}_-$) branches of the spectrum, respectively. Note that  the corresponding wavevectors are extrema of the particle-hole spectrum shown in Fig. \ref{spectrum}b. 
As a consequence, these two MF wavefunctions have different localization lengths in real space, $\xi^{(i)} = 1/k^{(i)}_{-} $ and $\xi^{(e)}=1/k^{(e)} $, which are inverse proportional to the corresponding gaps, $|\Delta_-|$ and $\Delta_{sc}$. Which one of them determines the localization length  of $\varPhi_{M}$  depends on the ratio between $\Delta_{Z}$ and $\Delta_{sc}$. 

In particular, as the magnetic field is being increased from zero to the critical value $B_{c}=2 \Delta_{sc}/g \mu_B$,  MFs  emerge at each end of the nanowire \cite{lutchyn_majorana_wire_2010,oreg_majorana_wire_2010,Sato}. However, if the localization lengths $\xi^{(i)}$ of these emerging MFs are comparable to the length $L_{sc}$ of the nanowire, then these two MFs are hybridized into a  subgap fermion of finite energy (see App. \ref{finite_wire}). This then  implies that the MFs in a finite wire can only appear  at a magnetic field 
$B_{c}^*\simeq  B_{c}(1+4\alpha_R/\Delta_{sc}L_{sc})$
that is {\it larger}  than the $B_c$ obtained for a semi-infinite nanowire. If the magnetic field is increased further, the 
main contribution to the MF bound state comes from the exterior branches.

\begin{figure}[tbp]
      \centering
      \includegraphics[width=8.5cm]{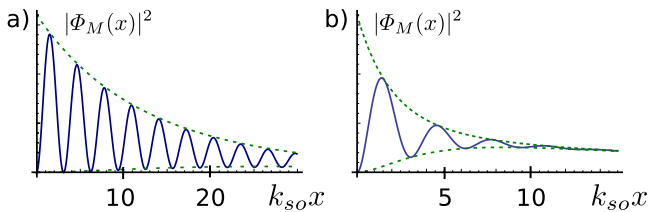}
\caption{ The MF probability density $|\varPhi_{M}(x)|^2$, see
Eq. (\ref{wave_MF}), for a nanowire in the strong SOI regime with  $\hbar^2
   \Delta_{sc}/m \alpha_R^2 =0.1$. The decaying MF wavefunction
$\varPhi_{M}$ oscillates with a period 
$\pi/k_{so}$.
(a) In the topological phase but still near the transition, $\Delta_Z = 2 \Delta_{sc}$,
the MF wavefunction undergoes many oscillations due to the
interference between $\widetilde{\varPhi}^{(i)}_-$ and $\widetilde{
\varPhi}^{(e)}_1$. (b) Deep inside the topological
phase, $\Delta_Z = 7 \Delta_{sc}$, the MF wavefunction from the interior
branches decays much faster than the one from the exterior branches.
This leads to only a few oscillations of the density and a uniform
decay with a decay wavevector $k^{(e)}$ away from the end of the
nanowire.}
\label{wave_SC_Strong}
\end{figure}

The composite structure of the MF wavefunction manifests itself in the probability density  
$|\varPhi_{M}(x)|^2$ along the nanowire. The density of a MF  
coming only from one of the branches, for example,  
$\widetilde{\varPhi}^{(i)}_-$, is just decaying exponentially. In  
contrast, the density of the composite MF exhibits oscillations  
(see Fig. \ref{wave_SC_Strong}). These oscillations  are due to interference and are most pronounced when the  
contributions of $\widetilde{ \varPhi}^{(e)}_1$ and  
$\widetilde{\varPhi}^{(i)}_-$ to $\widetilde{ \varPhi}_{M}$ are similar,  
{\it i.e.}, when both decay lengths,  $\xi^{(i)}$ and  
$\xi^{(e)}$, are close to each other.

The approach of the rotating magnetic field allows us to understand the structure of composite MF wavefunction. However, this approach is valid under the assumption that the SOI at the Fermi level is the largest energy scale. In order to explore the weak SOI regime, we come back to the full quadratic Hamiltonian in the next subsection.

\subsection{Weak SOI regime: Near the topological phase transition \label{sec:sc_weak}}

The regime of weak SOI is defined by the condition that the Zeeman splitting is much larger than the SOI energy at the Fermi level, $\Delta_Z \gg  m \alpha_R^2/ \hbar^2 $ [or $k_F\approx k_Z$, see Eq. (\ref{eq_kF})]. This allows us to treat the SOI as a perturbation.

Around the Fermi points, $\pm k_F$, the eigenstates of $\mathcal{H}^0$  are found from the Schr\"{o}dinger equation $\mathcal{H}^0(-i\partial_x\rightarrow\pm k_F) \varphi^{R/L} = 0$ [see Eq. (\ref{h_0_density})] and given by
\begin{equation}
\varphi^{R/L}_0 = \frac{1}{\sqrt{2}}\begin{pmatrix}
          -1 \pm \frac{ k_{so} }{k_F}\\1 \pm \frac{ k_{so} }{k_F}
         \end{pmatrix},
\label{Mag_Wave}
\end{equation}
where $\varphi^{R/L}_0$ denotes the eigenstates at $k=\pm k_F$. In Eq. (\ref{Mag_Wave}) we kept only terms up to first order in  $k_{so} / k_F$. As expected, $\varphi^R_0$ and $\varphi^L_0$ are nearly `aligned' along the magnetic field since $\Delta_Z \gg \alpha_R k_F$. In the absence of SOI, $\varphi^R_0$ and $\varphi^L_0$ are perfectly aligned along the $x$-axis and have the same spin, so they cannot be coupled by an ordinary $s$-wave superconductor. The SOI slightly tilts the spins in the orthogonal direction, which then allows the coupling between these states if the nanowire is brought into the proximity of an $s$-wave superconductor. 

The exterior branches can be treated in the linearized approximation similar to Sec. \ref{section_MF_SC_strong},
\begin{equation}
\chi (x) = R (x) e^{i k_{F} x}+ L (x) e^{-i k_{F} x},
\label{operator_1}
\end{equation}
where, again,  $R$ ($L$) annihilates  a right (left) moving electron. These operators are connected to spin-up ($\Psi_\uparrow$) and spin-down ($\Psi_\downarrow$) electron operators as $R (x) = \varphi^R (x) \cdot \psi$ and $L(x)=\varphi^L (x) \cdot \psi$, where $\psi=(\Psi_\uparrow,\Psi_\downarrow)$ (with corresponding support for right and left movers).
Here, $\varphi^{R/L} (x)$ is given by Eq. (\ref{Mag_Wave}) but where we allow now also for a slowly
varying x-dependence.

In this approximation, we find
\begin{align}
H^{0}
=-i \hbar \upsilon_F \int  dx \ [ R^\dagger (x)\partial_x R (x)-\ L^\dagger (x)\partial_x L (x)],
\end{align}
where the Fermi velocity is given by $\upsilon_F \approx \hbar \sqrt{2 \Delta_Z /m }$. The proximity-induced superconductivity [see Eq. (\ref{hm_sc})]  is described in the linearized model as
\begin{align}
H^{sc}=\frac{1}{2}\int dx\ \oline{\Delta}_{sc} (R (x)L (x)-L (x)R (x) +h.c.),
\label{hm_sc_w}
\end{align}
where  the strength of the proximity-induced effective $p$-wave superconductivity, $\oline{\Delta}_{sc} $, is found from Eqs. (\ref{hm_sc}), (\ref{Mag_Wave}), and (\ref{operator_1}), 
\begin{equation}
\frac{\oline{\Delta}_{sc} }{\Delta_{sc}}=
(\varphi^R_0)^*\cdot i \sigma_2 \varphi^L_0
 =2 \frac{k_{so} }{k_F}=\frac{\sqrt{2 m}\alpha_R}{\hbar \sqrt{\Delta_Z}}  \ll 1.
\label{gap_sc}
\end{equation}
The suppression of  $\oline{\Delta}_{sc} $ compared to $\Delta_{sc}$ can be understood from the fact that two states with opposite momenta at the Fermi level have mostly parallel spins due to the strong magnetic field and they slightly deviate in the orthogonal direction due to the weak SOI, which then leads to a suppression of $\Delta_{sc}$ by a factor $k_{so} / k_F$.

Again, introducing a vector $\phi^{(e)}(x) = (R, L, R^\dagger, L^\dagger)$, we represent the linearized Hamiltonian ${H}^{(e)}={H}^{0}+{H}^{sc}$ as
\begin{align}
{H}^{(e)}& =\frac{1}{2}\int dx (\phi^{(e)})^\dagger \mathcal{{H}}^{(e)} \phi^{(e)}, \nonumber \\
\mathcal{{H}}^{(e)}& = -i \hbar \upsilon_F \tau_3 \partial_x  +\oline{\Delta}_{sc}\tau_2 \eta_2,
\label{ham_d3}
\end{align}
where the Pauli matrices $\tau_{1,2,3}$ act on the right/left-mover subspace. 

The spectrum around the Fermi points in infinite space (no boundary conditions) follows from the Schr\"{o}dinger equation, $\mathcal{{H}}^{(e)} \varphi^{(e)} = E^{(e)} \varphi^{(e)} $,  and is given by $E^{(e)}_{1,2}=\pm\sqrt{(\hbar\upsilon_F \kappa)^2+\Delta_{e}^2}$, where the momentum $\kappa$ is  again taken from the Fermi points. Here, $2 \Delta_{e} \equiv 2 \oline{\Delta}_{sc} $ is the gap induced by superconductivity.

The zero-energy solutions that are normalizable for $x>0$ are two evanescent modes with  wavevector $\bar{k}^{(e)}=\oline{\Delta}_{sc} /\hbar \upsilon_F$ 
determining the localization length. These solutions can be written explicitly as $\varphi^{(e)}_1 = (1, -i, 1, i)e^{-\bar{k}^{(e)} x}$ and $\varphi^{(e)}_2 = (-i, 1, i, 1)e^{-\bar{k}^{(e)} x}$. Repeating the procedure that led us to Eq. (\ref{Gamma_oper}) in Sec. \ref{section_MF_SC_strong}, we can introduce a new MF operator,
\begin{equation}
\gamma = \int dx\ \varPhi_{E=0}(x)\cdot \Psi (x).
\label{Gamma_oper_1}
\end{equation}
Here $\Psi=(\Psi_\uparrow,\Psi_\downarrow, \Psi_\uparrow^\dagger,\Psi_\downarrow^\dagger)$.
The two corresponding  wavefunctions are written as 
\begin{align}
&\varPhi^{(e)}_j = \begin{pmatrix}
                   f_j(x)\\if_j^*(x)\\f_j^*(x)\\-if_j(x)
                  \end{pmatrix}e^{- \bar{k}^{(e)} x},
\label{weak_ext1}
\end{align}
where
\begin{align}
&f_1(x) =  i (1+k_{so}/k_F)e^{i k_F x}-(1-k_{so}/k_F)e^{-i k_F x},\nonumber\\
&f_2(x) = i (1-k_{so}/k_F)e^{-i k_F x} - (1+k_{so}/k_F)e^{i k_F x}
\label{weak_ext2}.
\end{align}

The effect of the SOI on the states around $k=0$ is negligible near the topological phase transition if $\hbar^2 |\Delta_-|/2m\alpha_R^2\ll1$. Therefore, the eigenstates for weak and strong SOIs are the same in first order in SOI. This means that we are allowed to take $\widetilde{\varPhi}^{(i)}_-$ and  $\widetilde{\varPhi}^{(i)}_+$ given by Eq. (\ref{vectors_inter_1}) and transform them back into the lab frame,
\begin{align}
\varPhi^{(i)}_-=\begin{pmatrix}
-i\ {\rm sgn} (\Delta_{-})\\
1\\
i\ {\rm sgn} (\Delta_{-})\\
1
\end{pmatrix} e^{-k_-^{(i)}x},\ 
\varPhi^{(i)}_+=\begin{pmatrix}
 1\\
-i\\
 1\\
i
\end{pmatrix} e^{-k_+^{(i)} x}.\label{weak_int}
\end{align}

After we found four basis wavefunctions $\{\varPhi^{(i)}_-, \varPhi^{(i)}_+, \varPhi^{(e)}_1, \varPhi^{(e)}_2\}$, we should  impose the boundary conditions on their linear combination $\varPhi(x)$. The wavefunction $\varPhi(x)$ should vanish at the boundary $x=0$. One can see that if we neglect the corrections to the wavefunctions $\varPhi^{(e)}_1$ and  $\varPhi^{(e)}_2$ coming from SOI, then we are able to satisfy the boundary conditions. This is a consequence of the fact that in the absence of SOI both states at the Fermi level have the same spin, so that the functions $\varPhi^{(e)}_i$,  $i=1,2$, effectively become  spinless objects and the MF always exists and arises only from the exterior branches. For the complete treatment, however, we should also consider contributions from the interior branches. This will be addressed next.

The set of wavefunctions $\{\varPhi^{(i)}_-, \varPhi^{(i)}_+, \varPhi^{(e)}_1, \varPhi^{(e)}_2 \}$ becomes linearly dependent in the topological regime $\Delta_{-}<0$ and the MF wavefunction is given by
\begin{equation}
\varPhi_{M}=\left(1- \frac{k_{so}}{k_F}\right)\varPhi^{(e)}_1 - \left (1- \frac{k_{so}}{k_F}\right)\varPhi^{(e)}_2 - 4 \frac{k_{so}}{k_F} \varPhi^{(i)}_-.
\label{MF_strong}
\end{equation}

As in the regime of strong SOI (see Subsec. \ref{section_MF_SC_strong}), the MF wavefunction has its support around wavevectors $k=0$ (interior branches) and  $k=\pm k_F$ (exterior branches). However, in contrast to the previous case, the contribution of the interior branches is suppressed by the small parameter $k_{so}/k_F$, thus the exterior branches contribute most to the MF wavefunction. At the same time we note that the localization length of MFs is determined by the smallest gap in the system. Near the topological phase transition\cite{lutchyn_majorana_wire_2010,oreg_majorana_wire_2010}, which corresponds to the closing of the topological (interior) gap, the interior branches determine the localization length as long as $\bar{k}^{(e)}>k_-^{(i)}$. If the magnetic field is increased further, the gap in the system is given by the exterior gap, $2\Delta_e=2\oline{\Delta}_{sc} \propto 1/\sqrt{B}$ [see Eq. (\ref{gap_sc})]. The localization length  is increasing as $\propto \sqrt{B}$. As soon as it is 
comparable to the nanowire 
length $L_{sc}
 $, the wavefunction of the two 
MFs at opposite ends overlap, and the two zero-energy MF levels are split into one subgap fermion of finite energy.

\begin{figure}[tbp]
       \centering
       \includegraphics[width=8.5cm]{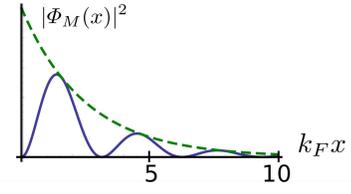}
\caption{ The MF probability density $|\varPhi_{M}(x)|^2$ for a nanowire in the weak SOI regime (  
$m \alpha_R^2 / \hbar^2 \Delta_{sc}=0.2$ and $\Delta_Z = 5 \Delta_{sc}$) oscillates with period $\pi/k_{F}$ due to 
interference between right- and left-moving contributions [see Eq. (\ref{weak_ext1})]. The decay length is given by $1/\bar{k}^{(e)}$.}
\label{wave_SC_Weak}
\end{figure}

In the weak SOI regime and sufficiently far away from the topological transition point, $\Delta_Z > \Delta_{sc} (1+k_{so}/k_{F})$, so that the gap is determined by the exterior branches only, we can work in the simplified model\cite{suhas_majorana} given by $\mathcal{{H}}^{(e)}$ [see Eq. (\ref{ham_d3})]. 
The explicit MF wavefunction can be found from Eq. (\ref{MF_strong}),
\begin{equation}\varPhi_{M}(x)=
\begin{pmatrix}
e^{-i\pi/4}\\
i e^{i\pi/4}\\
e^{i\pi/4}\\
-i e^{-i\pi/4}
\end{pmatrix} \sin (k_F x) e^{-\bar{k}^{(e)} x}.
\label{MF_weak_SOI}
\end{equation}
Again, we note that this wavefunction describes a 
MF with the spin of both, the electron and the hole, pointing in
 x-direction, again, up to corrections of order of $k_{so}/k_F$. The MF probability density $|\varPhi_{M}(x)|^2\propto\sin^2 (k_F x) e^{-2\bar{k}^{(e)} x}$ decays oscillating with a period 
 half the Fermi wavelength, $\lambda_F/2=\pi/k_F$ (see Fig. \ref{wave_SC_Weak}).

In passing we remark that $\mathcal{{H}}^{(e)}$ given in Eq. (\ref{ham_d3})   belongs to the topological class $\rm D \rom{3}$ according to the classification scheme of Ref. \onlinecite{Ryu_topological_class} and supports MFs in 1D,
in agreement with our result Eq. (\ref{MF_weak_SOI})
\cite{footnote_time_reversal}.

\section{Majorana fermions in NS junctions \label{section_normal_region}}

\begin{figure}[!bp]
    \centering
    \includegraphics[width=8cm]{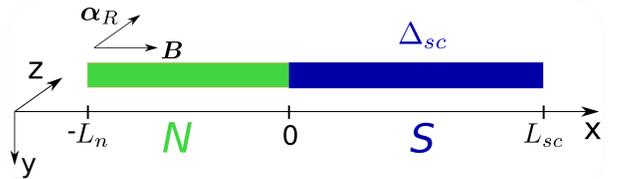}
\caption{NS junction of a nanowire.  The right section (blue) of the nanowire from $x=0$ to $x=L_{sc}$ is brought into contact with a bulk $s$-wave superconductor  (not shown) 
that induces a gap $\Delta_{sc}$ in the nanowire via proximity effect.
The left section (green) of the nanowire from $x=-L_{n}$ to $x=0$ is in the normal regime.
A magnetic field $\bf B$ is applied along the entire nanowire in $x$-direction and  perpendicularly to the Rashba SOI vector $\boldsymbol{\alpha}_R$, which points in $z$-direction.
}
\label{wire_normal}
\end{figure}

In this section we consider a nanowire containing a normal-superconducting (NS) junction where  the right part is in the superconducting and the left part in the normal regime, see
Fig. \ref{wire_normal}. The junction is assumed to be fully transparent.
We will show that the MF wavefunction leaks out of the superconducting section and
leads to a new MF bound state that extends over the entire normal section. We note that this bound state is different from Andreev bound states \cite{Andreev} known to occur
in NS junction systems. Indeed, the existence of the latter at zero energy would be accidental in the presence of a magnetic field since they move away from the Fermi level if the magnetic field is varied. Further, the MFs found in this section always exist in the topological regime and are not sensitive to the length $L_{n}$ of the normal section, in stark contrast to Andreev bound states that move in energy as function of $L_{n}$ \cite{Andreev,Riedel}.

We continue to work with the formalism developed in Sec. \ref{section_MF_SC} [see Eqs. (\ref{Gamma_oper}) and (\ref{Gamma_oper_1})] and represent  $\gamma$ in the basis of electron/hole spin-up/spin-down operators $\Psi (x)$ in terms of a four-component vector $\varPhi(x)$ on which we impose the boundary conditions. As before, the length of the superconducting part of the nanowire $L_{sc}$ is assumed to be much larger than any decay length given by $k_-^{(i)}$, $k_+^{(i)}$, or $k^{(e)}$ ($\bar{k}^{(e)}$).
This assumption allows us to treat the nanowire again as semi-infinite with no boundary conditions at $x=L_{sc}$. In contrast, the normal section, $x\in[-L_{n}, 0]$, is finite. 
Thus, at $x=0$ we invoke continuity of the wavefunctions and their derivatives and at $x=-L_{n}$ we impose vanishing boundary conditions,
\begin{align}
\varPhi (x=0^-) &= \varPhi (x=0^+),\label{boundary_1} \\
\partial_x \varPhi (x=0^-) &= \partial_x \varPhi(x=0^+),\\
\varPhi (x=-L_{n})&=0. 
\label{boundary_3}
\end{align}

The analytical form of the functions $\varPhi(x)$ can be found in two regimes, again in the weak and strong SOI limits, which we address now in turn.

\subsection{NS junction in the strong SOI regime} 

As before, the most convenient way to treat the strong SOI regime is to work in the rotating frame. In Sec. \ref{section_MF_SC_strong} we already found the four basis wavefunctions at zero energy
\begin{equation}
(\widetilde{\varPhi}^{(sc)}_1, \widetilde{\varPhi}^{(sc)}_2, \widetilde{\varPhi}^{(sc)}_3, \widetilde{\varPhi}^{(sc)}_4)=(\widetilde{\varPhi}^{(i)}_-, \widetilde{\varPhi}^{(i)}_+, \widetilde{ \varPhi}^{(e)}_1, \widetilde{ \varPhi}^{(e)}_2)
\label{sc_definitions}
\end{equation}
in the superconducting section, for $x\geq 0$  [see Eqs. (\ref{vectors_inter_1}) and (\ref{vectors_inter_2})].

The eigenfunctions for the normal section can be found from the linearized Hamiltonians for the interior branches, $\mathcal{\widetilde{H}}^{(i)}$ [see Eq. (\ref{H_int})], and for the exterior branches, $\mathcal{\widetilde{H}}^{(e)}$ [see Eq. (\ref{H_ext})], with $\Delta_{sc}=0$. The exterior branches are not gapped leading to the four propagating modes (see Fig. \ref{Strong_SOI}b) with wavefunctions  given by
\begin{widetext}
\begin{equation}
\widetilde{\varPhi}^{(n)}_1 = \begin{pmatrix}
-i\ e^{ik_{so}x}\\e^{-i k_{so}x}\\i\ e^{-ik_{so}x}\\e^{i k_{so}x}
\end{pmatrix},\label{wave_n_1}\ \ 
\widetilde{\varPhi}^{(n)}_2 = \begin{pmatrix}
e^{i k_{so}x}\\
i\ e^{-i k_{so} x}\\e^{-i k_{so}x}\\-i\ e^{i k_{so} x}
\end{pmatrix},\ \ 
\widetilde{\varPhi}^{(n)}_3 = \begin{pmatrix}
i\ e^{ik_{so}x}\\
e^{-i k_{so}x}\\-i\ e^{-ik_{so}x}\\e^{i k_{so}x}
\end{pmatrix}
,\ \ 
\widetilde{\varPhi}^{(n)}_4 = \begin{pmatrix}
e^{i k_{so}x}\\-i\ e^{-i k_{so} x}\\e^{-i k_{so}x}\\
i\ e^{i k_{so} x}
\end{pmatrix},
\end{equation}
\end{widetext}
where we choose to represent the wavefunctions in form of MFs, guided by our expectation that the final solution is also a MF. 
The interior branches are gapped by the magnetic field (see Fig. \ref{Strong_SOI}b) and the four corresponding  wavefunctions describing evanescent modes are given by
\begin{widetext}
\begin{align}
&\widetilde{\varPhi}^{(n)}_5=\begin{pmatrix}
            -i\ e^{-i k_{so} x}\\e^{i k_{so} x }\\  i\ e^{i k_{so} x}\\e^{-i k_{so} x }
            \end{pmatrix} e ^{k^{(n)} x},\ \ \ \ \ \ \ \ \ \ \ \ \ \ \ \ \ 
\widetilde{\varPhi}^{(n)}_6=\begin{pmatrix}
            e^{-i k_{so} x}\\i\ e^{i k_{so} x }\\ e^{i k_{so} x}\\-i\ e^{-i k_{so} x }
            \end{pmatrix} e ^{k^{(n)} x},\\[5pt]
&\widetilde{\varPhi}^{(n)}_7=\begin{pmatrix}
            i\ e^{-i k_{so} x}\\e^{i k_{so} x }\\  -i\ e^{i k_{so} x}\\e^{-i k_{so} x }
            \end{pmatrix} e ^{-k^{(n)}(L_n+x)},\ \ \ \ \ \ \
\widetilde{\varPhi}^{(n)}_8=\begin{pmatrix}
           e^{-i k_{so} x}\\-i\ e^{i k_{so} x }\\  e^{i k_{so} x}\\i\ e^{-i k_{so} x }\\ 
            \end{pmatrix} e ^{-k^{(n)}(L_n+x)},\label{wave_n_8}
\end{align}
\end{widetext}
with  $k^{(n)} =\Delta_{Z}/\alpha_R$. The modes $\widetilde{\varPhi}^{(n)}_{5,6}$ decay from their maximum at $x=0$ to zero for $x\rightarrow -\infty$, while $\widetilde{\varPhi}^{(n)}_{7,8}$  decay from their maximum at $x=-L_n$ to zero for $x\rightarrow +\infty$.

After having introduced the basis consisting of 12 MF wavefunctions given by Eqs. (\ref{vectors_inter_1}), (\ref{vectors_inter_2}), (\ref{wave_n_1}), and (\ref{wave_n_8}), we search for  their linear combination,
\begin{equation}
\widetilde{\varPhi}_M (x) 
 = \left\{
  \begin{array}{l l}   
\sum_{j=1}^{4} a_j \widetilde{\varPhi}^{(sc)}_j, & \quad x\geq0 \vspace*{5 pt}\\
\sum_{j=1}^{8} b_j \widetilde{\varPhi}^{(n)}_j, & \quad -L_{n}\leq  x \leq0, 
  \end{array} \right.
\label{sum_MF}
\end{equation}
such that the boundary conditions (\ref{boundary_1})-(\ref{boundary_3}) are satisfied. This is, in general, possible only in the topological phase. However, we also find solutions in the nontopological phase, where these solutions exist only if some special relations between the parameters $\Delta_Z$, $\alpha_R$, $\Delta_{sc}$, and $L_n$ are satisfied. This allows us to identify them as  Andreev bound states in an NS junction. Since they are not  of interest here,  we focus on the solutions in the topological phase only. 

\begin{figure}[tbp]
        \centering
        \includegraphics[width=7.0cm]{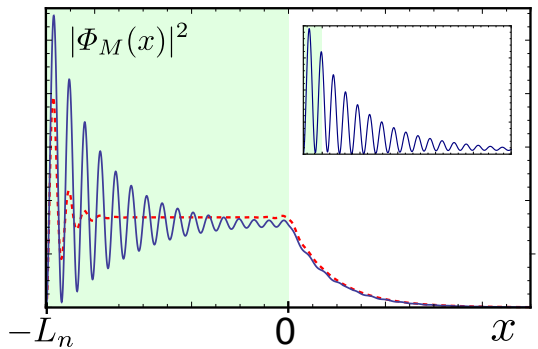}
\caption{The MF probability density $|\varPhi_{M}(x)|^2$ in an NS  
junction for a nanowire in the strong SOI regime ($\hbar^2  
\Delta_{sc}/m \alpha_R^2 =0.06$). The normal section of length $L_n$ is long compared
to the decay length, {\it i.e.} $k^{(n)}L_{n}\gg1$.
 The MF wavefunction extends over the  
entire normal section (green) and  decays  exponentially
inside the superconducting section (white). The oscillations with period $\pi/k_{so}$  result from interference between  
the three components $\widetilde{\varPhi}^{(n)}_{2,3,7}$,   
of $\varPhi_{M}$. In   
weak magnetic fields ($\Delta_Z = 1.5 \Delta_{sc}$, blue full line), $k^{(n)}L_{n}\sim 1$, and
the oscillations extend over the entire normal section. In contrast, in strong magnetic  
fields ($\Delta_Z = 7 \Delta_{sc}$, red dashed line),  the oscillations  
are strongly suppressed. Inset: $|\varPhi_{M}(x)|^2$ as function of $x$ for a short normal section, $k^{(n)}L_{n}\ll1$. Similarly to the case of the superconducting wire (see Fig. \ref{wave_SC_Strong}), $|\varPhi_{M}(x)|^2$  decays oscillating.}
\label{Fig:wave_SC_Strong_NS}
\end{figure}

It is worth of pointing out that due to the internal symmetry of the MF wavefunctions,  five of the coefficients are readily seen to vanish, namely
\begin{equation}
a_2=b_1=b_4=b_5=b_8=0.
\end{equation} 
The exact analytical solution is given in  App. \ref{app_exact_solution} and used for the plot in Fig. \ref{Fig:wave_SC_Strong_NS}. Here, we only discuss the two limiting cases of  long and short
normal sections $L_n$.

First, we consider $L_n\gg 1/k^{(n)}$, allowing us to neglect the terms 
$\widetilde{\varPhi}^{(n)}_5$ and $\widetilde{\varPhi}^{(n)}_6$ at $x=-L_n$, and 
$\widetilde{\varPhi}^{(n)}_7$ and $\widetilde{\varPhi}^{(n)}_8$ at $x=0$.
In this case, the sum  
in Eq. (\ref{sum_MF}) is determined by the coefficients
\begin{align}
&a_1  \rightarrow 0,\ \ \ \ b_6  \rightarrow 0, \  \ \ \ b_7=-1,\label{coef_strong_soi}\\
&a_3 =  b_3=\cos (2k_{so}L_n),\  \  \  \ a_4 =b_2=-\sin (2k_{so}L_n), \nonumber \\
\nonumber
\end{align}
leading to the solution of the form
\begin{widetext}
\begin{equation}
\widetilde{\varPhi}_M(x) = \left\{
  \begin{array}{l l}   
\widetilde{\varPhi}^{(n)}_3(2L_n+x)\ e^{-k^{(e)}x}\ ,& \quad x>0\\
\widetilde{\varPhi}^{(n)}_3(2L_n+x) - 
\widetilde{\varPhi}^{(n)}_3(-x)\ e^{-k^{(n)}(x+L_n)}\ ,& \quad -L_{n}\leq  x <0.
  \end{array} \right.
\end{equation}
\end{widetext}
Thus, we see that the MF wavefunction in the lab frame, ${\varPhi}_M(x)$,  decays monotonically in the superconducting section while it oscillates in the normal one (see Fig. \ref{Fig:wave_SC_Strong_NS}). 
In  weak magnetic fields the MF probability density $|{\varPhi}_M(x)|^2$ oscillates over the entire normal section, in contrast to the near absence of oscillations in strong magnetic fields. 

We note that a long normal section serves as a `momentum filter'. As shown in Subsec. \ref{section_MF_SC_strong}, a MF has equal support from the exterior and interior branches [see Eq. (\ref{wave_MF})] if the entire nanowire is in the superconducting regime. In contrast to that, if a significant portion of the nanowire is in the normal regime, the MF has support mostly from the exterior branches with momenta $k\simeq \pm k_F$. The contributions from the interior branches with momenta $k\simeq0$ are negligibly small, $a_1\rightarrow 0$ and $a_2=0$. This behavior can be understood in terms of momentum mismatch: 
the normal section does not have  propagating modes with $k=0$ (in the lab frame). Thus, while such $k=0$ modes exist in the superconducting section, they cannot propagate into the normal section.

Second, we consider the opposite limit $L_n\ll 1/k^{(n)}$. Here, we can treat the decaying solutions, $\widetilde{\varPhi}^{(n)}_j$, $j=5,6,7,8$, as being constant over 
$L_n$. The MF wavefunction $\widetilde{\varPhi}_M(x)$ is constructed from seven basis MF wavefunctions with the 
same coefficients as in Eq. (\ref{coef_strong_soi}) with the only difference that now $a_1=-1$.
For  short normal sections, the form of the MF probability density $|{\varPhi}_M(x)|^2$ is very similar to the one of a superconducting nanowire (compare  inset of Fig. \ref{Fig:wave_SC_Strong_NS} with Fig. \ref{wave_SC_Strong}a). Again, the interference between $\widetilde{\varPhi}^{(i)}_-$, $\widetilde{ \varPhi}^{(e)}_1$, and  $\widetilde{ \varPhi}^{(e)}_2$ leads to  oscillations in the superconducting section.

In both limits of short and long normal sections, $|{\varPhi}_M(x)|^2$ has it maximum in the normal section while it decays in the superconducting one. This opens the possibility of measuring the presence of a MF state spectroscopically in the normal section. The amplitude and period of oscillations of the MF probability density is sensitive to magnetic fields and to the nanowire length, controlled e.g. by an infinite barrier on the left end. Moreover, by shifting such a barrier via gates, we can change the type of  MF from $\widetilde{ \varPhi}^{(e)}_1$ to $ \widetilde{ \varPhi}^{(e)}_2$ [see Eqs. (\ref{sc_definitions}), (\ref{sum_MF}), and (\ref{coef_strong_soi})]. This amounts to change a given MF state from a `real part' type, $\psi +\psi^\dagger$, to an `imaginary part' type,
$i(\psi -\psi^\dagger)$.

\subsection{NS junction in the weak SOI regime}
As before, we first identify basis wavefunctions in the superconducting section and in the normal section. Then, we search for a linear combination of them such that the boundary conditions given by Eqs. (\ref{boundary_1})-(\ref{boundary_3}) are satisfied.

As shown in Sec. \ref{sec:sc_weak}, the MF wavefunction $\varPhi_{M}$ has predominantly support from the exterior branches. The correction to the MF wavefunction from the interior branches is suppressed by a factor $k_{so}/k_F$ [see Eq. (\ref{MF_strong})]. If we focus on the regime away from the topological phase transition where the exterior gap is smaller than the interior one, then, as in Eq. (\ref{weak_ext1}),  the MF wavefunction can be constructed to first order  in $k_{so}/k_F$ from the  exterior wavefunctions  $\varPhi^{(e)}_{j=1,2}$ 
alone,
\begin{align}
&\varPhi^{(sc)}_{j} = \begin{pmatrix}
                   g_j(x)\\ig_j^*(x)\\g_j^*(x)\\-ig_j(x)
                  \end{pmatrix}e^{-\bar{k}^{(e)} x}, \nonumber  \\[5pt]
&g_1=e^{-i\pi/4}  \sin (k_F x), \ \ \ g_2=e^{-i\pi/4}  \cos (k_F x).
\label{weak_ext_NS}
\end{align}

The propagating electron modes $\varphi^R$ and $\varphi^L$ of Eq. (\ref{Mag_Wave}) in the normal section described by $H^0$ (without $H^{sc}$) were considered before
[see Eq. (\ref{Gamma_oper_1})] and given by
\begin{widetext}
 \begin{equation}
 \varPhi^{(n)}_1 =\begin{pmatrix}
          e^{-ik_Fx} \\ -e^{-ik_Fx}\\e^{ik_Fx}\\-e^{ik_Fx}
         \end{pmatrix},\ \
 \varPhi^{(n)}_2 =\begin{pmatrix}
          e^{ik_Fx} \\ -e^{ik_Fx}\\e^{-ik_Fx}\\-e^{-ik_Fx}
         \end{pmatrix},\ \
 \varPhi^{(n)}_3 =\begin{pmatrix}
         i e^{-ik_Fx} \\ -i  e^{-ik_Fx}\\-i e^{ik_Fx}\\ i e^{ik_Fx}
         \end{pmatrix},\ \
 \varPhi^{(n)}_4 =\begin{pmatrix}
          i e^{ik_Fx} \\-i e^{ik_Fx}\\-i e^{-ik_Fx}\\ i e^{-ik_Fx}
         \end{pmatrix}.
 \end{equation}
\end{widetext}

The ansatz for the wavefunction in both sections is
\begin{equation}
\varPhi(x) = \left\{
  \begin{array}{l l}   
\sum_{j=1}^{2} a_j \varPhi^{(sc)}_j , & \quad x\geq0\vspace*{5 pt}\\
\sum_{j=1}^{4} b_j \varPhi^{(n)}_j ,& \quad -L_{n}\leq  x \leq0.
  \end{array} \right.
\label{sum_MF_NS}
\end{equation}
The  coefficients $a_j$ and $b_j$ can then be found from the boundary conditions (\ref{boundary_1})-(\ref{boundary_3}), 
\begin{eqnarray}
a_1&=&2[\bar{k}^{(e)} \sin(k_F L_n)+k_F \cos (k_F L_n)],\nonumber\\ 
a_2&=&2 k_F \sin (k_F L_n),\nonumber\\
b_1&=&-b_4=k_F \cos (k_F L_n - \pi/4),\nonumber\\
b_2&=&-b_3=-k_F \cos (k_F L_n + \pi/4),
\end{eqnarray}
leading finally to the MF wavefunction of the form
\begin{align}
\varPhi_M(x)=f(x) \begin{pmatrix}
e^{-i \pi/4}\\-e^{-i \pi/4}\\
e^{i \pi/4}\\-e^{i \pi/4}
\end{pmatrix},
\end{align}
where
\begin{equation}
f(x) = \left\{
  \begin{array}{l l}   
k_F \sin (k_F[x+L_n]), &  -L_{n}\leq  x \leq0 \vspace*{5 pt}\\
 e^{-\bar{k}^{e}x}[ k_F \sin (k_F[x+L_n])
 \\ +\bar{k}^{e} \sin( k_F x) \sin( k_F L_n) ], & x\geq0.
  \end{array} \right.
\label{FUN_mf}
\end{equation}

\begin{figure}[bp]
        \centering
        \includegraphics[width=6.5cm]{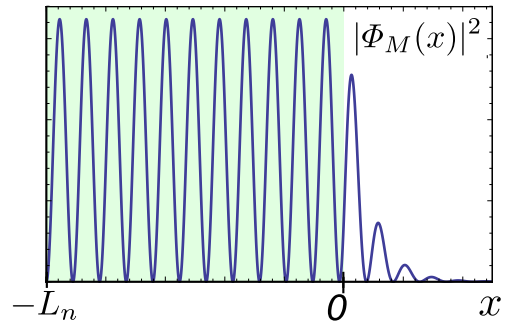}
\caption{ The MF probability density $|\varPhi_{M}(x)|^2$ in an NS  
junction for a nanowire in the weak SOI regime ($m \alpha_R^2/\hbar^2  
\Delta_{sc} =0.2$ and $\Delta_Z = 5 \Delta_{sc}$).  
The MF wavefunction extends over the entire normal section of length $L_n$ (green) and  decays rapidly
inside the superconducting section (white).}
\label{Fig:NS_weak}
\end{figure}

The corresponding MF probability density $|\varPhi_{M}(x)|^2$ is shown in Fig. \ref{Fig:NS_weak}. The MF wavefunction extends over the entire normal section. In this section, we considered as basis functions only propagating modes, $\varPhi^{(n)}_j$, leading to a purely oscillatory solution with the period given by half the Fermi wavelength $\lambda_F/2=\pi/k_F$. In contrast, in the superconducting section, the MF wavefunction decays on a short distance. In other words, the MF is mostly delocalized over the entire normal section and is strongly localized in the superconducting section, in agreement with recent numerical results\cite{Chevallier}. This might simplify the detection of MFs by local density measurements since the normal section is freely accessible to tunnel contacts,
in contrast to the superconducting section which needs to be covered by a bulk $s$-wave superconductor.

\section{Conclusions \label{section_conclusions}} In this work, we have focused on the  wavefunction properties of Majorana fermions occurring in  superconducting nanowires and in nanowires with an NS junction. The superconducting phase is effectively  $p$-wave  and is based on an interplay of $s$-wave proximity effect, spin orbit interaction, and magnetic fields. 
We have derived explicit  results for the MF wavefunctions in the regime of strong and weak SOI and shown that the wavefunctions are composite objects, being superpositions 
of contributions coming from the interior (around $k=0$) and exterior (around $\pm k_F$)  branches of the spectrum in momentum space. While the underlying Hamiltonians considered in this work allow degenerate MF wavefunctions, the boundary conditions at hand completely lift this degeneracy and we are left with only one single MF state at a given end of the nanowire
({\it i.e.} in total there are two MF states for the entire nanowire).

In the strong SOI regime of a superconducting nanowire both branches contribute equally. However, the decay length of the MF is determined by the branch that also defines the smallest gap in the system. Moreover, the oscillations in the MF probability density with period of the Fermi wavelength  are decaying on the scale given by the largest gap in the system. 
In the weak SOI regime, the exterior branches mostly contribute to the MF wavefunction. 
The contributions of the interior branch is suppressed by the small factor $k_{so}/k_F\ll 1$ and only close to the topological phase transition this branch determines the localization length of the MF. The interference between modes from $k_F$ and $-k_F$ leads to  oscillations  of the  
 probability density of the MF with an exponentially decaying envelope.

For a nanowire with an NS junction we find that the MF wavefunction becomes delocalized over the entire normal section, while still being localized in the superconducting section, in agreement
with recent numerical results.\cite{Chevallier}
Again, we obtain analytical results for the weak and strong SOI regimes.
Depending on the length of the normal section, the support of the MF wavefunction is centered at zero momentum or at the Fermi points. Again,  we find different localization lengths and oscillation periods of the MF
in the normal section that are tunable by  magnetic fields. Based on this insight, we expect that in a tunneling density of states measurement the tunneling current at zero bias exhibits oscillations as a function of position along the normal section due to the presence of the MF in the normal section.

Finally, we remark that in this work we have focused on single particle properties and ignored, in particular, interaction effects. It would be interesting to extend the present analysis to interacting Luttinger liquids, in particular for the SOI nanowire with an NS junction, combining
the approaches developed in Refs. \onlinecite{Maslov,braunecker,suhas_majorana,stoudenmire}.

\acknowledgments
We acknowledge fruitful discussions with Karsten Flensberg, Diego Rainis, and Luka Trifunovic.
This work is supported by the Swiss NSF, NCCR Nanoscience and NCCR QSIT, and DARPA.

\vspace{15 pt}

\appendix

\section{Finite nanowire \label{finite_wire}}

In this Appendix we address the problem of a finite superconducting section of length $L_{sc}$. In this case, the decaying modes 
$\widetilde{\varPhi}_{0}=\{\widetilde{\varPhi}^{(i)}_-,\ \widetilde{\varPhi}^{(i)}_+,\ \widetilde{ \varPhi}^{(e)}_1,\ \widetilde{ \varPhi}^{(e)}_2\}$
with maximum at $x=0$ are given by Eqs. (\ref{vectors_inter_1}) and (\ref{vectors_inter_2}). Now we  should also take into account the four evanescent modes with maximum at $x=L_{sc}$. These modes,
$\widetilde{\Phi}_{L_{sc}}=\{\widetilde{\Phi}^{(i)}_-,\ \widetilde{\Phi}^{(i)}_+,\ \widetilde{ \Phi}^{(e)}_1,\ \widetilde{ \Phi}^{(e)}_2\}$,
 are found from Eqs. (\ref{H_int}),  (\ref{H_ext}) and are similar by their structure to $\widetilde{\varPhi}_{0}$,

\begin{widetext}
\begin{equation}
\begin{array}{l l}
 \widetilde{\Phi}^{(i)}_-=\begin{pmatrix}
i\ {\rm sgn} (\Delta_{-}) e^{-ik_{so} x}\\
e^{ik_{so} x}\\
-i\ {\rm sgn} (\Delta_{-})e^{ik_{so} x}\\
e^{-ik_{so} x}
\end{pmatrix} e^{k_-^{(i)}(x-L_{sc})}, \quad \quad & \quad \quad
\widetilde{\Phi}^{(i)}_+=\begin{pmatrix}
 e^{-ik_{so} x}\\
i\ e^{ik_{so} x}\\
 e^{ik_{so} x}\\
-i\ e^{-ik_{so} x}
\end{pmatrix} e^{-k_+^{(i)} (x-L_{sc})},\vspace{10 pt}\\ 
\widetilde{ \Phi}^{(e)}_1=\begin{pmatrix}
-i\ e^{ik_{so} x}\\
e^{-ik_{so} x}\\
i\ e^{-ik_{so} x}\\
e^{ik_{so} x}
\end{pmatrix} e^{-k^{(e)} (x-L_{sc})}, \quad \quad  & \quad \quad 
\widetilde{ \Phi}^{(e)}_2=\begin{pmatrix}
 e^{ik_{so} x}\\
-i\ e^{-ik_{so} x}\\
 e^{-ik_{so} x}\\
i\ e^{ik_{so} x}
\end{pmatrix} e^{-k^{(e)}(x-L_{sc})}.
\end{array}
\label{vectors_inter_L}
\end{equation}
\end{widetext}

We construct a $8\times 4$ matrix $\widetilde{\omega}(x)$ from the eight basis wavefunctions. The zero-energy solution can then be compactly written as
\begin{equation}
\widetilde{\varPhi}(x) = {\bf{a}}\cdot \widetilde{\varPhi}_{0}(x) + {\bf{b}}\cdot\widetilde{\Phi}_{L_{sc}}(x) \equiv\widetilde{\omega}(x) \begin{pmatrix}
                                                                                                                          {\bf{a}}\\  {\bf{b}}
                                                                                                                         \end{pmatrix}\, ,
\end{equation}
where ${\bf{a}}=(a_1, a_2, a_3, a_4)$ and ${\bf{b}}=(b_1, b_2, b_3, b_4)$ are coefficients that should be determined from the boundary conditions,
\begin{eqnarray}
&\widetilde{\varPhi}(x=0, L_{sc})=0,
\end{eqnarray}
which can be rewritten as a matrix equation in terms of a $8\times8$ matrix $\widetilde{\varOmega}$,
\begin{equation}
\widetilde{\varOmega} \begin{pmatrix}
 {\bf{a}}\\ {\bf{b}}
\end{pmatrix}=\begin{pmatrix}
\widetilde{\omega}(0)\
\widetilde{\omega}(L_{sc})
\end{pmatrix} 
\begin{pmatrix}
 {\bf{a}}\\ {\bf{b}}
\end{pmatrix}=0.
\end{equation}

The determinant of the matrix $\widetilde{\varOmega}$ is nonzero, so the solution of the matrix equation is unique and trivial, $({\bf{a}}, {\bf{b}})=0$. This means that, strictly speaking,  MFs cannot emerge in a finite-size nanowire.  MFs exist only under the assumption that the overlap of the two MF wavefunctions (localized at each end of the nanowire and derived in a semi-infinite nanowire model) can be neglected. Otherwise, the two MFs are hybridized into a subgap fermion of finite energy.

\section{Exact solution in strong SOI regime \label{app_exact_solution}}

Here, we present the exact solution for the MF wavefunction $\widetilde{\Phi}(x)$ composed of seven different basis MFs wavefunctions [see Eq. (\ref{sum_MF})] and satisfing the boundary conditions  (\ref{boundary_1})-(\ref{boundary_3}). We find
 \begin{widetext}
 
\begin{align}
&a_2=b_1=b_4=b_5=b_8=0,\nonumber\\
&a_1 = 4 k_{so}^2  + 4 k_{so} k^{(e)} \cosh (k^{(n)} L) \sin (2k_{so}L_n)
-2 k^{(e)}\cos (2k_{so}L_n) [k^{(e)} \cosh (k^{(n)} L_n) + k^{(n)} \sinh (k^{(n)} L_n)],\nonumber\\
&a_3 = e^{-k^{(n)} L_n} [ e^{k^{(n)} L_n} (k^{(e)}[k^{(n)}-2  k^{(i)}_-]+k^{(n)}[k^{(n)}-k^{(i)}_-])-4 k_{so}^2 e^{2 k^{(n)} L_n} \cos(2k_{so}L_n)\nonumber\\
&-2 k_{so} (e^{2 k^{(n)} L_n}k^{(e)}+k^{(i)}_--k^{(n)})\sin(2k_{so}L_n)],\nonumber\\
&a_4 = - e^{-k^{(n)} L_n}[2 e^{k^{(n)} L_n} k_{so}(k^{(n)}-k^{(i)}_--k^{(e)}) + 2 k_{so} (e^{2k^{(n)} L_n}k^{(e)} +k^{(i)}_--k^{(n)}) \cos(2k_{so}L_n)\nonumber\\
& -  e^{2 k^{(n)} L_n}4 k_{so}^2\sin(2k_{so}L_n)],\nonumber\\
&b_2 =- e^{-k^{(n)} L_n}(-2 e^{k^{(n)} L_n}k_{so}k^{(e)}+2k_{so}(k^{(i)}_--k^{(n)} ) \cos(2k_{so}L_n)-4 k_{so}^2e^{2 k^{(n)} L_n}\sin(2k_{so}L_n)),\nonumber\\
&b_3=e^{-k^{(n)} L_n}(e^{k^{(n)} L_n}k^{(e)} [k^{(e)}-k^{(i)}_-]-4 k_{so}^2e^{2 k^{(n)} L_n} \cos(2k_{so}L_n)+2 k_{so} (k^{(n)}-k^{(i)}_-)\sin(2k_{so}L_n)),\nonumber\\
&b_6=-2 k_{so}(k^{(n)}-k^{(i)}_-)-e^{k^{(n)} L_n} k^{(e)}[2  k_{so}\cos(2k_{so}L_n)+(k^{(e)}-k^{(i)}_-)\sin(2k_{so}L_n)],\nonumber\\
&b_7 = 4 k_{so}^2 e^{k^{(n)} L_n}.
\end{align}
\end{widetext}

\end{document}